\documentclass[a4paper,fleqn]{cas-sc}
\usepackage{soul}
\usepackage[sort&compress,numbers]{natbib}
\usepackage{setspace}
\def\tsc#1{\csdef{#1}{\textsc{\lowercase{#1}}\xspace}}
\tsc{WGM}
\tsc{QE}
\tsc{EP}
\tsc{PMS}
\tsc{BEC}
\tsc{DE}
\begin{document}
\let\WriteBookmarks\relax
\def\floatpagepagefraction{1}
\def\textpagefraction{.001}
\shorttitle{Mechanical and Thermal Stability of Free-standing Monolayer Amorphous Carbon}
\shortauthors{Levi C. Felix et~al.}
%\begin{frontmatter}

\title [mode = title]{On the Mechanical Properties and Thermal Stability of a Recently Synthesized Monolayer Amorphous Carbon}                
\author[1,2]{Levi C. Felix}
\author[1,2]{Raphael M. Tromer}
\author[3]{Luiz A. Ribeiro Junior}
\author[1,2]{Douglas S. Galvao}[orcid=0000-0003-0145-8358]
\cormark[1]
\ead{galvao@ifi.unicamp.br}

\address[1]{Applied Physics Department, 'Gleb Wataghin' Institute of Physics, State University of Campinas, Campinas,SP, 13083-970, Brazil}
\address[2]{Center for Computing in Engineering \& Sciences, State University of Campinas, Campinas, SP, 13083-970, Brazil}
\address[3]{Institute of Physics, University of Bras\'ilia , 70919-970, Bras\'ilia, Brazil}

\cortext[cor1]{Corresponding author}

\begin{abstract}
Recently (C.-T. Toh \textit{et al.}, Nature \textbf{577}, 199 (2020)), the first synthesis of free-standing monolayer amorphous carbon (MAC) was achieved. MAC is a pure carbon structure composed of five, six, seven and eight atom rings randomly distributed. MAC proved to be surprisingly stable and highly fracture resistant. Its electronic properties are similar to boron nitride. In this work, we have investigated the mechanical properties and thermal stability of MAC models using fully-atomistic reactive molecular dynamics simulations. For comparison purposes the results are contrasted against pristine graphene (PG) models of similar dimensions. Our results show that MAC and PG exhibit distinct mechanical behavior and fracture dynamics patterns. While PG after a critical strain threshold goes directly from elastic to brittle regimes, MAC shows different elastic stages between these two regimes. Remarkably, MAC is thermally stable up to 3600 K, which is close to the PG melting point. These exceptional physical properties make MAC-based materials promising candidates for new technologies, such as flexible electronics. 
\end{abstract}

%\includegraphics[scale=0.08]{figs/graphical-abstract.png}

%\item Monolayer amorphous carbon has a melting point in $\sim 3600$ K;  
%\item Monolayer amorphous carbon present different elastic stages during the fracture process;
%\item The formation of linear atomic carbon chains prior to structural rupture under tensile loading is more pronounced in monolayer amorphous carbon than in pristine graphene;

%\linenumbers


\maketitle
\doublespacing

\section{Introduction}
Graphene-based nanomaterials have emerged in the past decade as an ideal platform for developing new flexible electronics with a good cost-efficiency relationship \cite{chang_AFM,pumera_EES,pumera_CSR,perreault_CSR}. Recent advances have broadened their application framework ranging from stretchable circuits to flexible energy storage devices \cite{secor_AM,cheng_NATCOMM,gwon_EES}. The success achieved by graphene has inspired several studies to obtain other classes of carbon allotropes in manufacturing new solutions for optoelectronics  \cite{zhang_FP,rajkamal_AMT}. Recently, the first free-standing monolayer of amorphous carbon (MAC) was reported \cite{toh_2020}. MAC is an amorphous material that contains sp$^2$ and sp$^3$-like hybridized carbon atoms forming a network with a wide distribution of bond length bond angle values, and containing five- to up eight-member rings \cite{toh_2020}. Importantly, short-range order exists in MAC, but the inter-atomic distances and inter-bonding angles are different regarding pristine graphene (PG) lattices.

Rapid progress in different synthetic routes for amorphous carbon allotropes has been recently obtained \cite{falcao_JCTB,joo_SciAdv,toh_2020}. Zachariasen carbon monolayer, a new amorphous 2D carbon allotrope with one-atom-thick, was synthesized on the germanium surface by Joo and coworkers \cite{joo_SciAdv}. In their study, an in-plane fully sp2-hybridized carbon network was obtained at high temperatures (>900$^{\circ}$C). Very recently, laser-assisted chemical vapor deposition was employed to synthesize free-standing, continuous, and stable MAC, with lattice arrangement different from disordered graphene \cite{toh_2020}. The results have shown that the MAC deforms to a high breaking strength, without crack propagation from the point of fracture. In the obtained MAC structure, the ring distribution does not resemble a Zachariasen-like random network. Further investigations on MAC thermal stability and mechanical properties are needed to provide a more detailed understanding of this remarkable material that can pave the way for its possible applications in flexible electronics.

Herein, inspired by the recent MAC synthesis \cite{toh_2020}, the mechanical and thermal stability of a model MAC lattice was investigated in the framework of fully-atomistic reactive molecular dynamics simulations. The MAC mechanical behavior was investigated using the stress-strain relationship and fracture toughness. We also investigated MAC thermal stability. For comparison purposes, we have performed the same set of simulations for PG models of the same size dimensions

\section{Materials and Methods}
\begin{figure}[pos=ht]
	\centering
		\includegraphics[width=\linewidth]{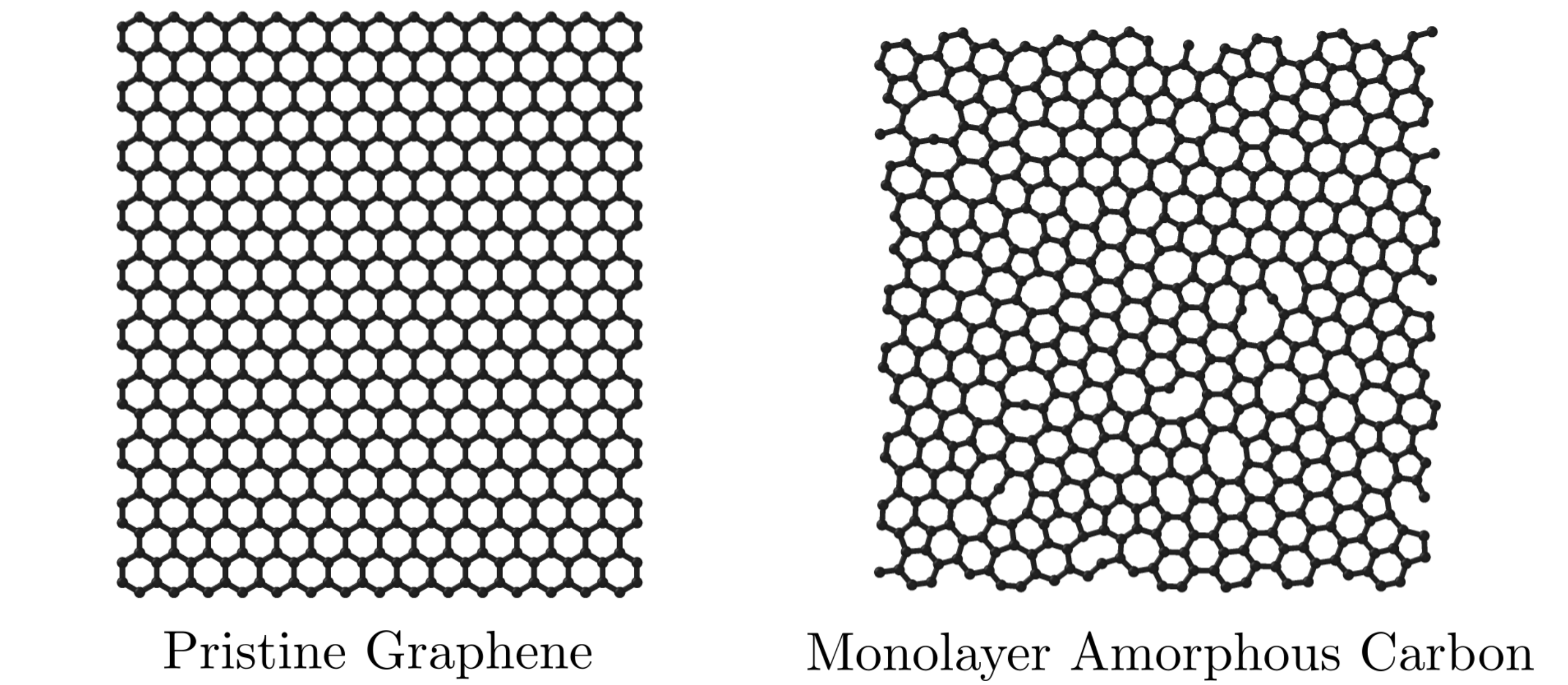}
	\caption{Atomic structure of the model lattices: (left) pristine graphene and (right) monolayer amorphous carbon.}
	\label{fig:structures}
\end{figure}

We have carried out fully-atomistic molecular dynamics (MD) simulations using the adaptive-interatomic reactive bond-order (AIREBO) \cite{stuart_2000} potential as implemented by the large-scale atomic/molecular massively parallel simulator (LAMMPS) \cite{plimpton_1995}. In Figure \ref{fig:structures} we present the corresponding PG and MAC models. We used the same atomistic structural MAC model reported in its synthesis paper \cite{toh_2020}, which contains 610 carbon atoms. The PG model of similar dimensions contains 640 atoms, which is denser than the MAC one because MAC is a porous structure.

The tensile stretching simulation was performed  by increasing (up to$100\%$ of their initial length size) the cell dimensions. Finite boundary conditions were imposed on all of the simulation box walls. Before applying the uniaxial stretching, a thermalization within an NVT ensemble was performed at $300$ K for 100 ps to eliminate any residual stresses. Subsequently, a group of atoms at both ends of the structures were moved linearly in time to simulate an engineering strain rate of $10^{-6}$ fs$^{-1}$. From the virial stress tensor and its correspondence with continuum media \cite{subramaniyan_2008}, we estimated the elastic properties (such as the Young modulus) and the von Mises stress \cite{mises_1913} per atom $k$, defined as:
\begin{equation}
    \sigma^{k}_{v} = \sqrt{\frac{(\sigma^{k}_{xx} - \sigma^{k}_{yy})^2 + (\sigma^{k}_{yy} - \sigma^{k}_{zz})^2 + (\sigma^{k}_{xx} - \sigma^{k}_{zz})^2 + 6((\sigma^k_{xy})^2+(\sigma^k_{yz})^2+(\sigma^k_{zx})^2)}{2}},
\end{equation}
where $\sigma^k_{xx}$, $\sigma^k_{yy}$, and $\sigma^k_{zz}$ are the components of the normal stress and $\sigma^k_{xy}$, $\sigma^k_{yz}$, and $\sigma^k_{zx}$ are the components of the shear stress. The von Mises stress values provide useful information on the fracture dynamics from the starting stretching regime up to complete structural failure (fracture).

The thermal stability of PG and MAC models were investigated by heating both structures from $100$ K up to $10000$ K under periodic boundary conditions. Before heating, a thermalization/equilibration within an NPT ensemble was carried out to eliminate any residual thermal stresses and adjust the simulation box dimensions along the plane of the structures. The heating process is simulated by a linear temperature increase (thermal ramp) during $1$ ns. The free software OVITO \cite{stukowski_2010} was used to visualize MD snapshots and trajectories.

\section{Results and Discussions}
\begin{figure}[pos=ht]
	\centering
		\includegraphics[width=\linewidth]{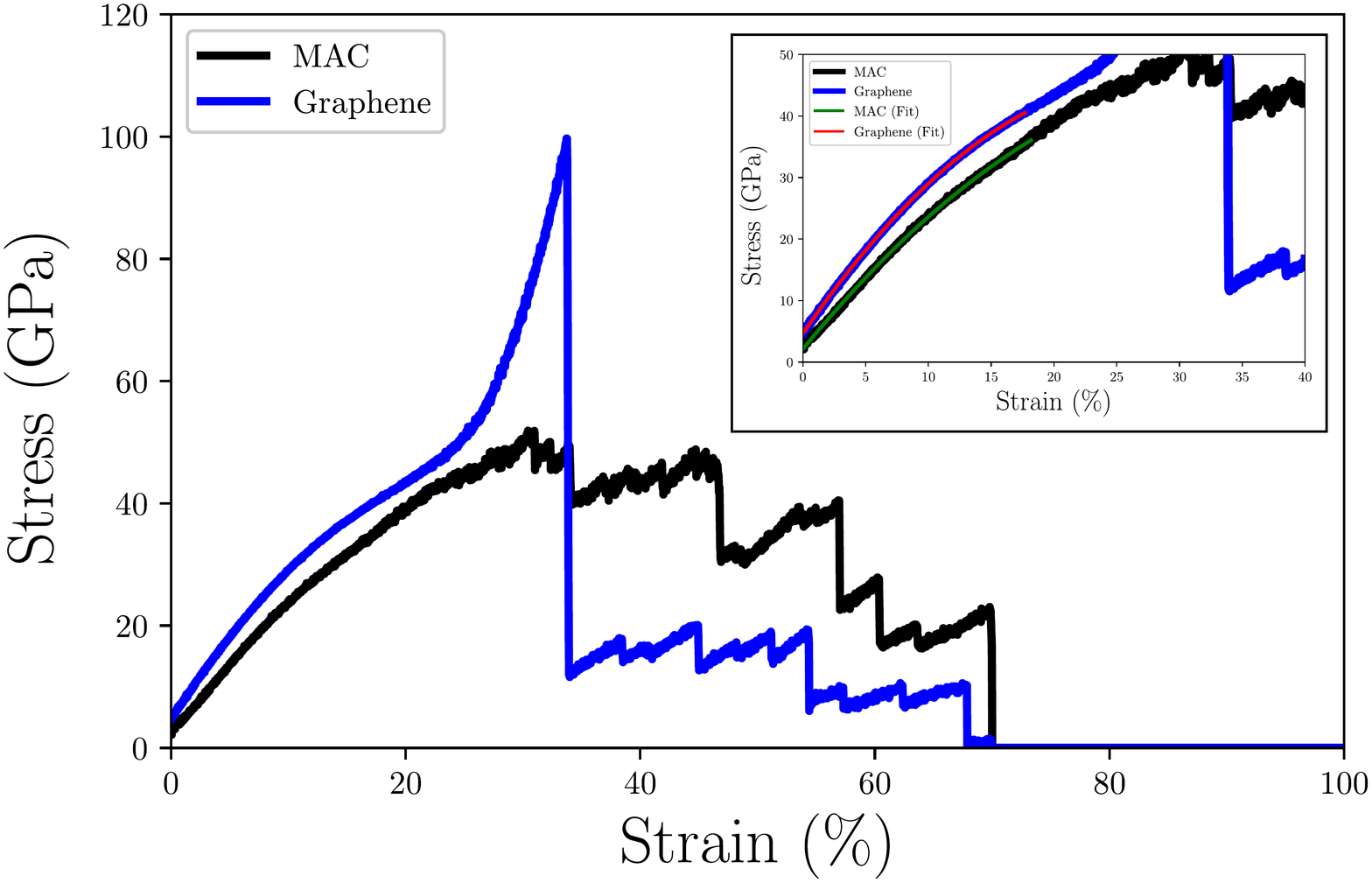}
	\caption{Stress-strain curves for both MAC (black) and PG (blues). The inset panel illustrates the fitting curves used to derive the parameters presented in Table \ref{table1}.}
	\label{fig:stress-strain}
\end{figure}

\begin{figure}[pos=ht]
	\centering
		\includegraphics[width=\linewidth]{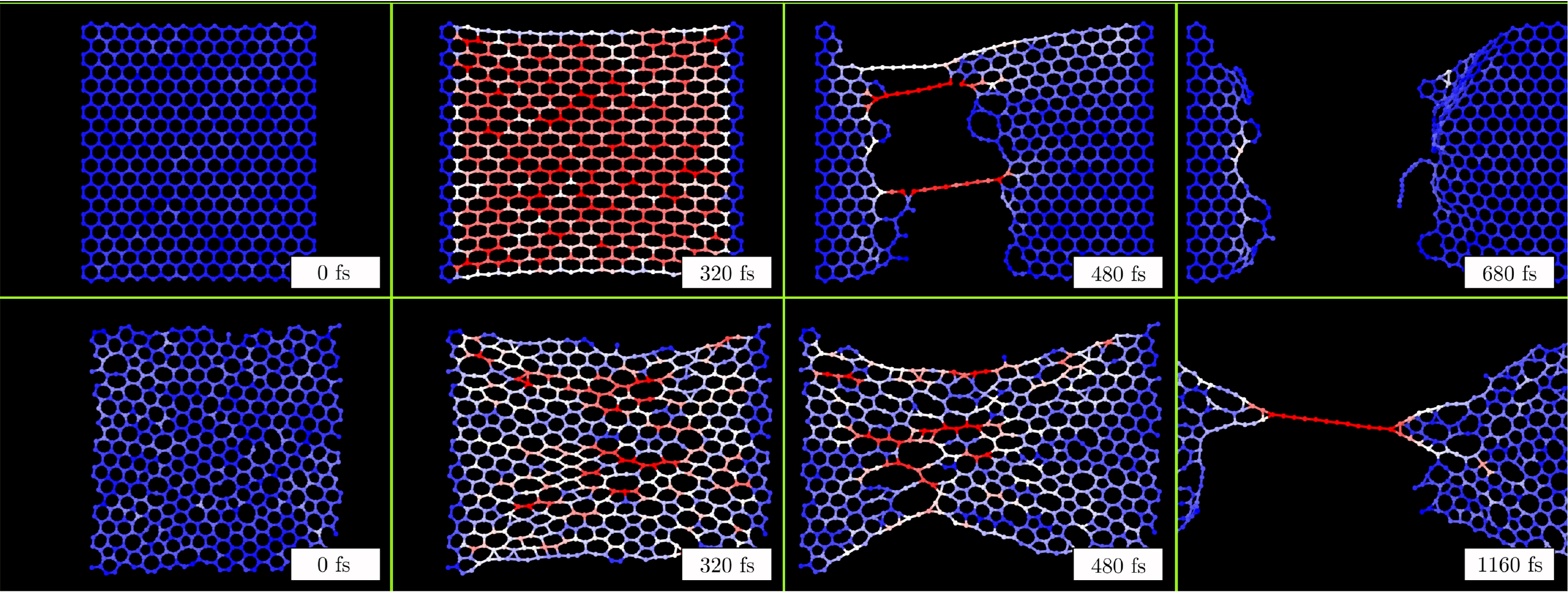}
	\caption{Representative MD snapshots from the uniaxial tensile stretching simulation for PG (top sequence of panels) and MAC (bottom sequence of panels). In the color scheme, blue and red stand for low and high values of von Mises stress, respectively.}
	\label{fig:mechanical-snapshots}
\end{figure}

In Figure \ref{fig:stress-strain} we present the calculated stress-strain curves for PG and MAC and in Figure \ref{fig:mechanical-snapshots} we present representative MD snapshots of the strain simulations. It is shown four different strain stages for PG (top) and MAC (bottom), respectively. Blue and red refer to low and high von Mises stress values, respectively. We can see from these Figures that PG and MAC present distinct elastic behavior and fracture patterns. 

In our simulations, the membranes were stretched at 300 K at a constant rate until total rupture. Previous graphene works have described the low-strain region as being quadratic due to the strong nonlinear behavior in the elastic regime \cite{lee_2008,zandiatashbar_2014,zhong_2019}. In this context, the stress-strain relationship is given by $\sigma = E\varepsilon + D\varepsilon^2$, where $\sigma$, $\varepsilon$, $E$, and $D$ are the symmetric second Piola-Kirchhoff stress, uniaxial strain, Young's modulus, and third-order elastic modulus, respectively \cite{thurston_1964}. The stress-strain curves shown in Figure \ref{fig:stress-strain} presents three common and distinct regions: (I) a quadratic elastic regime is observed up to a maximum stress value (the tensile strength ($TS$)), (II) then the stress values decrease when the structures start to fracture until they ultimately break (III) where the stress values goes abruptly to zero. 

The stress-strain curve for graphene shows an inflection point just after the elastic regime at $\sim 20\%$ strain, which is in agreement with other MD simulations using AIREBO \cite{zhong_2019}. The rapid oscillations between the elastic region and the fracture strain regimes are due to the formation of linear atomic carbon chains (LAC) that can be observed on the MD snapshots, shown in Figure \ref{fig:stress-strain}, which will be discussed later. The appearance of LAC has also been reported on previous experimental works \cite{jin_2009,chuvilin_2009} of tensile deformation of free-standing graphene membranes. 

PG presents practically one stage of elastic deformation, followed by an abrupt fracture, which is characteristic of a brittle behavior. MAC, on the other hand, exhibits a more complex fracture dynamics. It does not exhibit an abrupt fracture as observed in PG, but several stages where the stress drops associated with significant structural reconstructions followed by LAC formation in the final fracture stages. These sequential MAC reconstructions into metastable configurations are possible due to its large distribution of bond lengths and bond angles values, and the presence of different types of rings, which make possible several degrees of internal rearrangements and multiple stress release channels. 

The main significant differences between PG and MC in terms of fracture mechanisms can be explained by the distinct local concentration of von Mises stress throughout the atomic structure. For PG the von Mises stress is more uniformly distributed/accumulated and for high values (in red in Figure \ref{fig:mechanical-snapshots}) it results in an abrupt rupture and fast crack propagation. It is important to remark that in the simulations the strain is applied only along one of the two orthogonal directions (x and y) of the unit cell. When PG is stretched along the x-direction, the potential energy gains are stored into the C--C bonds that are almost parallel to this direction. These bonds are the first to break, and the PG is completely fractured at 480 fs (simulation time). Other results from MD simulations in the literature, concerning carbon-based 2D materials with long-range crystalline order, have shown that different fracture process might be obtained when the stretching is applied to distinct directions \cite{han_CMS,rakib_PB,JIAO_PLA,Verma_2018}. Since MAC is an amorphous material without long-range periodicity, we expect no impact of the stretching direction on its fracture mechanisms. For MAC the high-stress values are concentrated at localized regions causing a fracture to occur on just some regions of the structure and allowing structural reconstructions, which results that MAC complete structural failure occurs later (1160 fs) than PG. Also, LAC reconstructions are more pronounced on MAC with the existence of several line defects. The whole processes can be better understood from videos 1 and 2 in the Supplementary Materials.

The inset in Figure \ref{fig:stress-strain} shows the fitting curves (green for MAC and red for PG) from which the elastic properties presented in Table \ref{table1} were obtained. In Table \ref{table1} we present Young's modulus, third-order elastic modulus ($D$), tensile strength ($TS$) and fracture strain ($\varepsilon_{F}$) values for PG and MAC. Both $E$ and $D$ are obtained from fitting the quadratic regions of the stress-strain curves presented in the inset of Figure \ref{fig:stress-strain}. It can be seen that the values of $E$ for both materials are of the same order of magnitude with MAC being softer due to the presence of structural disorder. Also, the negative values obtained for $D$ indicate a decrease in stiffness as the strain increases, which in turn leads to an intrinsic strength\cite{lee_2008}. The strain value at the moment where the fracture starts defines $\varepsilon_F$ and $TS$ corresponds to the stress value at the fracture strain. Although MAC and PG start to break at almost the same strain value, the $TS$ value for graphene is almost twice that of MAC. The mechanical properties obtained here for the PG case are in good agreement with previous theoretical values reported in the literature using Tersoff \cite{Rajasekaran_Tersoff}, AIREBO \cite{zhong_2019,anthea_AIREBO}, and ReaxFF \cite{jensen_REAX} potentials.

\begin{table}[width=.95\linewidth,cols=5,pos=ht]
\caption{Mechanical Properties of MAC and PG: Young's modulus ($E$), third-order elastic modulus ($D$), fracture strain ($\varepsilon_{F}$) and Tensile Strength ($TS$) values.}
\begin{tabular*}{\tblwidth}{@{} LLLLL@{} }
\toprule
\textbf{Structure} & \boldmath$E$ \textbf{[GPa]} & \boldmath$D$ \textbf{[GPa]} & \boldmath$\varepsilon_{F}$ \boldmath$[\%]$ & \boldmath$TS$ \textbf{[GPa]} \\
\midrule
MAC & 249.14 & -3.43 & 30.43 & 51.90 \\
Pristine Graphene & 296.77 & -5.32 & 33.75 & 99.69 \\
\bottomrule
\end{tabular*}
\label{table1}
\end{table}

\begin{figure}[pos=ht]
	\centering
		\includegraphics[scale=.8]{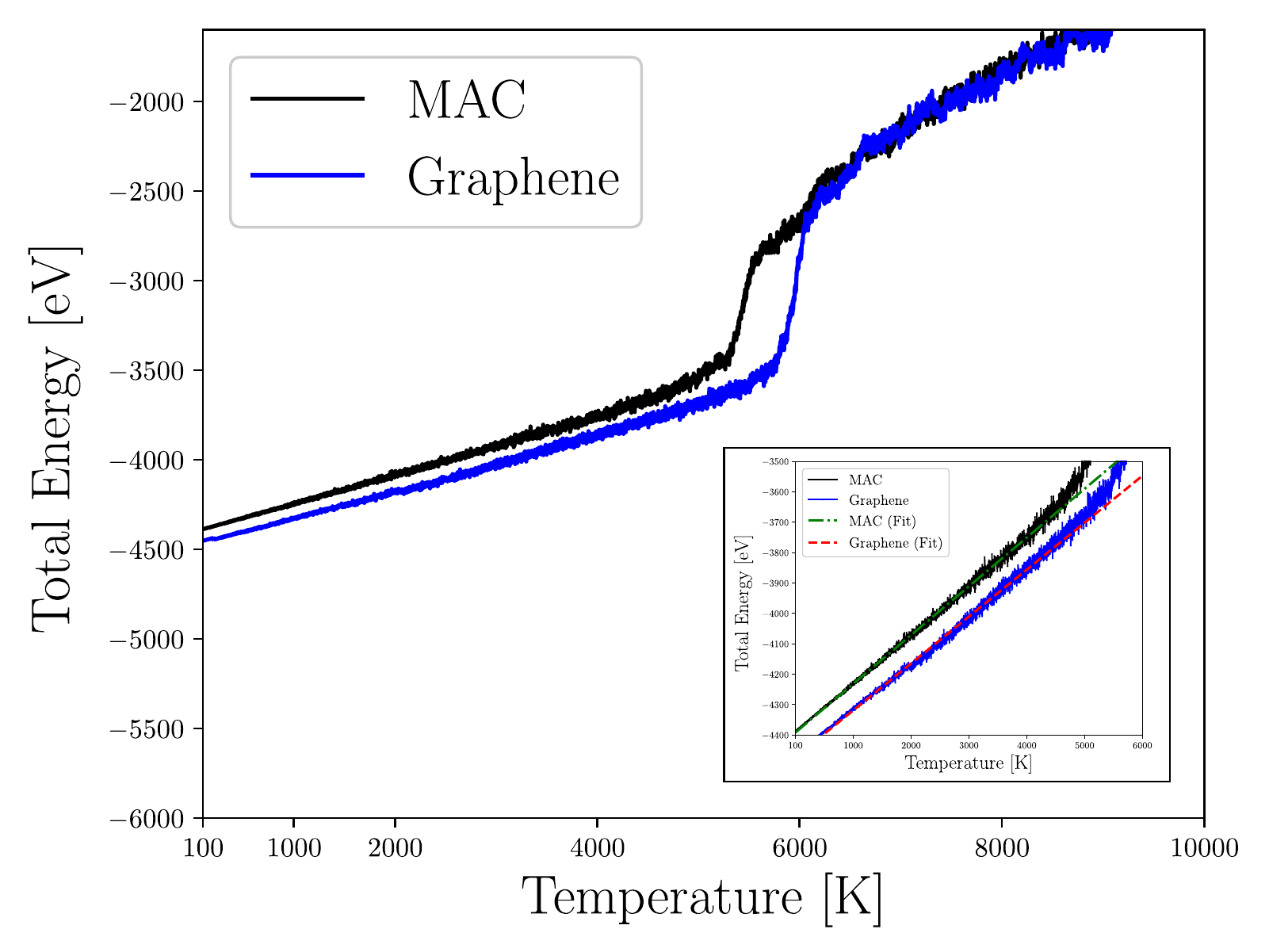}
	\caption{Total energy as a function of temperature for both MAC (black) and PG (blues). The fitting curves illustrated in the inset panel were used to estimate the sublimation temperatures.}
	\label{fig:temp-ramp}
\end{figure}

In Figure \ref{fig:temp-ramp} we present the results for the temperature dependence of the total energy for PG (blue curve) and MAC (black curve), respectively. The MD simulations were carried out using heating ramp protocols, with temperature varying from 100 up to 10000 K. This final temperature is substantially above the graphene melting point (or sublimation temperature), which is approximately 4510 K \cite{xia_ACSNANO,FOMIN_CARBON,singh_PRB} and it was used to warrant the almost complete evaporation of both structures.

We can see from Figure \ref{fig:temp-ramp} that there are two temperature regimes in which the total energy increases linearly with the two curves almost parallel and MAC values always slightly larger than the corresponding PG ones. The significant discontinuity presented by both curves marks a phase transition from a solid to gas-like structures. For the structures considered here the temperatures for these transitions occur at $4095$ K and $3626$ K, for PG and MAC, respectively (see Figure S1 in the supplementary materials). Previous works\cite{los_2015,ganz_2017,fomin_2020} have predicted a melting point for graphene between $4000$ and $6000$ K through atomistic simulations. The fitting curves illustrated in the inset panel of Figure \ref{fig:temp-ramp} were used to derive these sublimation temperatures (see Figure S2 in the supplementary materials). The critical temperature threshold for the PG case obtained here agrees with other results from MD simulations reported in the literature \cite{xia_ACSNANO,FOMIN_CARBON}. Surprisingly, the melting line for MAC is very close to the PG one, which is strong evidence of the remarkable thermal stability of the new carbon amorphous phase. Above these critical temperatures, there is a sharp increase in total energy indicating a completely molten structure. The abrupt increase in energy and the higher slope degrees for the curves in Figure \ref{fig:temp-ramp} are related to the higher atom velocities in the gas-like phase, which leads the system to accumulate a considerable amount of kinetic energy at high temperatures. Moreover, a substantial part of the harmonic and torsional energies associated with bond length and angles in the solid phase is converted into kinetic energy, increasing the amount of such a contribution for total energy (see Figure S3 in the Supplementary Materials).

\begin{figure}[pos=ht]
	\centering
		\includegraphics[width=\linewidth]{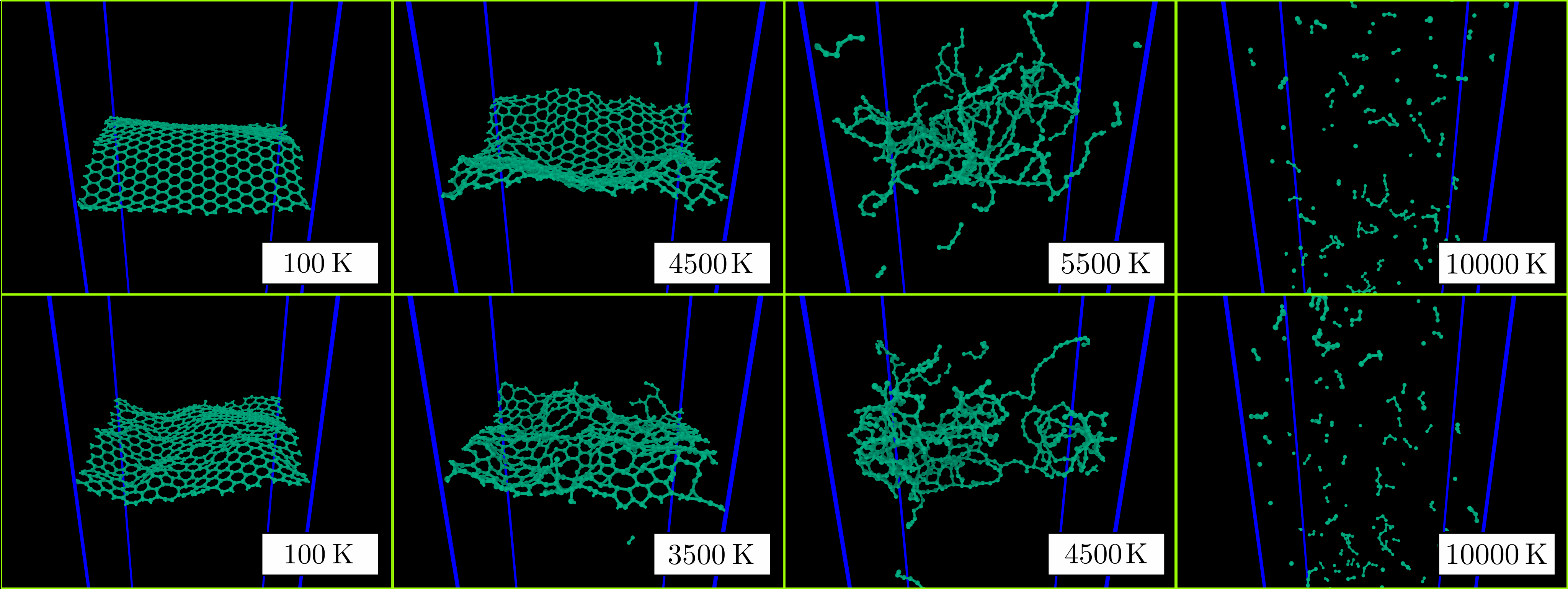}
	\caption{Representative MD snapshots of the melting simulations for PG (top sequence of panels) and MAC (bottom sequence of panels).}
	\label{fig:thermal-snapshots}
\end{figure}

In Figure \ref{fig:thermal-snapshots} we present representative MD snapshots of the melting simulations for PG (top) and MAC (bottom), respectively. The blue lines in these panels indicate the simulation box edges. The flat morphology of PG and MAC membranes favors the formation of LAC random coil conformations during the melting process, as can be seen in the panels for 5500 K (PG) and 4500 K (MAC), respectively. At elevated temperatures, the thermal fluctuations lead the sheets to wrinkle and fracture. This wrinkling effect also favors LAC formation. Above $\sim$ 6000 K for both structures, isolated atoms and small chain fragments (connecting 2-10 atoms) are observed. Therefore, the final morphologies for PG and MAC melting simulations resemble a gas-like phase of carbon atoms. In fact, a recent work \cite{fomin_2020} using MD simulations with AIREBO potential has shown that sublimation occurs instead of melting for graphene at high temperatures. The whole process can be better visualized in videos 3 and 4 of the Supplementary Materials.

\section{Summary and Conclusions}

We have investigated through fully atomistic reactive molecular dynamics simulations, the mechanical behavior and thermal stability of a recently synthesized free-standing monolayer amorphous carbon structure (MAC) \cite{toh_2020}. MAC is a pure carbon structure composed of randomly distributed 5,6,7 and 8 carbon rings. For comparison purposes, we have also considered pristine graphene (PG) of similar dimensions.

Our results show that the stress-strain curves for PG and MAC are quite distinct. While PG presents practically one stage of elastic deformation, followed by an abrupt fracture MAC does not exhibit an abrupt fracture but several stages where the stress value drops. These stages are associated with significant structural
reconstructions followed by linear atomic chain (LAC) formation in the final fracture stages. These structural reconstructions into
metastable configurations are possible due to MAC large distribution of bond lengths and bond angles values, and the presence of different types of rings. These features make possible several degrees of internal atomic rearrangements and multiple
stress release channels. The fracture dynamics is also different, while PG presents abrupt fractures and fast crack propagation, MAC exhibits localized fractures allowing structural reconstructions that prevents fast crack propagation and works as efficient stress release mechanisms. MAC elastic properties indicate a softer material than PG, which is expected due to the presence of structural disorder (amorphous). For instance, the Young's modulus is 249.14 GPa, in comparison to 296.77 GPa for PG.  With relation to MAC thermal stability, it presents remarkable stability with a melting line very close to graphene. For the structures considered here $4095$ K and $3626$ K, for PG and MAC, respectively. We hope the present work will stimulate further studies on this new carbon structure.

\section*{Acknowledgements}
The authors gratefully acknowledge the financial support from Brazilian research agencies CNPq, FAPESP, and FAP-DF. L.A.R.J acknowledges the financial support from a Brazilian Research Council FAP-DF and CNPq grants $00193.0000248/2019-32$ and $302236/2018-0$, respectively. DSG thanks the Center for Computing in Engineering and Sciences at Unicamp for financial support through the FAPESP/CEPID Grants \#2013/08293-7 and \#2018/11352-7.

%\section*{Supplementary Material}
%\noindent This electronic supplementary information presents additional data concerning statistical analysis for the linear fitting performed on the data presented in Figure \ref{fig:temp-ramp} in the main manuscript. The melting temperature is defined as the one corresponding to the timestep where the increase in total energy starts to deviate from linear behaviour, shown in Figure S1. The measurement of the latter deviation is based on the determination coefficient $R^2$. The maximum value of $R^2$ shown inn Figure S2 determines where the deviation from the linear behavior occurs.

%\begin{figure}[pos=ht]
%	\centering
%		\includegraphics[scale=.9]{figs/fit-thermal.png}
%	\caption{Figure S1. Total energy as a function of MD timesteps with the performed linear fit for both pristine graphene and MAC.}
%	\label{fig:fit-thermal}
%\end{figure}

%\begin{figure}[pos=ht]
%	\centering
%		\includegraphics[scale=.9]{figs/determination-thermal.png}
%	\caption{Figure S2. $R^2$ coefficient as a function of strain to determine where the increase in energy deviates from linear behavior with temperature.}
%	\label{fig:determination-thermal}
%\end{figure}

%Appendix sections are coded under \verb+\appendix+.

%\verb+\printcredits+ command is used after appendix sections to list 
%author credit taxonomy contribution roles tagged using \verb+\credit+ 
%in frontmatter.

\bibliographystyle{unsrt}
%\bibliographystyle{cas-model2-names}

\bibliography{cas-refs}

\end{document}